\documentclass[aps, prd, preprint, superscriptaddress, showpacs, nofootinbib]{revtex4}
\usepackage{graphicx,epstopdf}
\usepackage{ulem}
\usepackage{epsfig}
\usepackage{color}
\usepackage{multirow}
\usepackage{epstopdf}
\definecolor{My_red}        {cmyk}{0.00, 1.00, 1.00, 0.20}

\newcommand{\bmat}{\left(\begin{array}}
\newcommand{\emat}{\end{array}\right)}
\newcommand{\beq}{\begin{equation}}
\newcommand{\eeq}{\end{equation}}

\newcommand{\lsim}{\mathrel{\ltap}}

\usepackage[centertags]{amsmath}
\usepackage{amssymb}

\newcommand{\missingET}{\mathchoice{\rlap{\kern.2em/}E_T}{\rlap{\kern.2em/}E_T}{\rlap{\kern.1em$\scriptstyle/$}E_T}{\rlap{\kern.1em$\scriptscriptstyle/$}E_T}}

\let\jnfont=\rm
\def\NPB#1,{{\jnfont Nucl.\ Phys.\ B }{\bf #1},}
\def\PLB#1,{{\jnfont Phys.\ Lett.\ B }{\bf #1},}
\def\EPJC#1,{{\jnfont Eur.\ Phys.\ Jour.\ C }{\bf #1},}
\def\PRD#1,{{\jnfont Phys.\ Rev.\ D }{\bf #1},}
\def\PRL#1,{{\jnfont Phys.\ Rev.\ Lett.\ }{\bf #1},}
\def\MPLA#1,{{\jnfont Mod.\ Phys.\ Lett.\ A }{\bf #1},}
\def\JPG#1,{{\jnfont J.\ Phys.\ G }{\bf #1},}
\def\CTP#1,{{\jnfont Commun.\ Theor.\ Phys.\ }{\bf #1},}
\def\JHEP#1,{{\jnfont JHEP \ }{\bf #1},}
\def\NPPS#1,{{\jnfont Nucl.\ Phys.\ Proc.\ Suppl.\ }{\bf #1},}
\def\CPC#1,{{\jnfont Computl.\ Phys.\ Commun.\ }{\bf #1},}
\def\CPL#1,{{\jnfont Chin.\ Phys.\ Lett. }{\bf #1},}
\def\APPB#1,{{\jnfont Acta\ Phys.\ Polon.\ B }{\bf #1},}

\def\lsim{\raise0.3ex\hbox{$<$\kern-0.75em\raise-1.1ex\hbox{$\sim$}}}
\def\gsim{\raise0.3ex\hbox{$>$\kern-0.75em\raise-1.1ex\hbox{$\sim$}}}

\begin{document}

\title{An explicit calculation of pseudo-goldstino mass at the leading three-loop level}

\author{Jianpeng Dai}
\affiliation{CAS Key Laboratory of Theoretical Physics, Institute of Theoretical Physics,
Chinese Academy of Sciences, Beijing 100190, China}
\affiliation{School of Physics, University of Chinese Academy of Sciences, Beijing 100049, China}

\author{Tao Liu}
\affiliation{Institute of High Energy Physics, Chinese Academy of Sciences, Beijing 100049, China}
\affiliation{School of Physics, University of Chinese Academy of Sciences, Beijing 100049, China}

\author{Jin Min Yang}
\affiliation{CAS Key Laboratory of Theoretical Physics, Institute of Theoretical Physics,
Chinese Academy of Sciences, Beijing 100190, China}
\affiliation{School of Physics, University of Chinese Academy of Sciences, Beijing 100049, China}

\begin{abstract}
Pseudo-goldstinos appear in the scenario of multi-sector SUSY breaking.
Unlike the true goldstino which is massless and absorbed by the gravitino,
pseudo-goldstinos could obtain mass from radiative effects.
In this note, working in the scenario of two-sector SUSY breaking with gauge mediation,
we explicitly calculate the pseudo-goldstino mass
at the leading three-loop level and provide the analytical results 
after performing Taylor expansions in the loop integrals.
In our calculation we consider the general case of messenger masses (not necessarily equal)
and include the higher order terms of SUSY breaking scales.
Our results can reproduce the numerical value estimated  previously at the leading order of
SUSY breaking scales with the assumption of equal messenger masses. It turns out that the
results are very sensitive to the ratio of messenger masses, while the higher order terms
of SUSY breaking scales are rather small in magnitude. Depending on the ratio of messenger
masses, the pseudo-goldstino mass can be as low as ${{\cal O} (0.1)}$ GeV.

\end{abstract}

\maketitle

\section{Introduction}
Albeit lack of direct evidence at the LHC, supersymmetry (SUSY) as a generalization of space-time symmetry
in quantum field theory remains one of the most popular extensions of the Standard Model of particle physics.
The SUSY breaking and mediation mechanism plays a key role in the phenomenology of SUSY.
If the spontaneous SUSY breaking is generated independently in different hidden sectors, then each sector
provides a massless goldstino $\eta_i$ with SUSY breaking scale $F_i$. One linear combination of $\eta_i$
is identified as the true massless goldstino that is absorbed by gravitino, other orthogonal combinations
named pseudo-goldstinos could acquire masses from radiative corrections.
However, unconstrained by the supercurrent, the interactions between pseudo-goldstinos and
ordinary superparticles could be sizable enough to induce unconventional signatures in collider
physics and cosmology \cite{Argurio:2011hs,Cheung:2010mc,Cheung:2010qf,Craig:2010yf,McCullough:2010wf,Cheng:2010mw,Izawa:2011hi,Thaler:2011me,Cheung:2011jq,Dudas:2011kt,Argurio:2011gu,Liu:2013sx,Ferretti:2013wya,Hikasa:2014yra,Liu:2014lda,Cao:2020oxq,Chen:2021omv}.
For example, these physical pseudo-goldstinos with loop induced masses could be lighter than the
lightest neutralino so that the neutralino could decay to a pseudo-goldstino  plus a gauge boson
or Higgs boson, leading to intriguing phenomenology at the LHC and future lepton colliders
\cite{Hikasa:2014yra,Liu:2014lda,Chen:2021omv}.
On the other hand, since the thermally produced bino-like neutralinos may decay readily to pseudo-goldstinos
(the pseudo-goldstinos finally decay to gravitino dark matter),  the stringent constraints
on SUSY from dark matter detection and relic density can be relaxed (note that most GUT-constrained SUSY models
like CMSSM/mSUGRA suffer from such a tension between
dark matter constraints and muon g-2 explanation if the dark matter is the lightest bino-like neutralino
\cite{Wang:2021bcx}).

Obviously, the mass of pseudo-goldstino is a crucial parameter for all the phenomenological
analysis in the multi-sector SUSY breaking scenario.
Due to the intrinsic property of supergravity, there is a universal mass which is twice the
gravitino mass $m_{3/2}$ at tree level. Another tree-level contribution coming from electroweak symmetry
breaking in the low energy part is negligible. The possible loop corrections to the mass of pseudo-goldstino
have been discussed in Ref.~\cite{Argurio:2011hs}. If SUSY breaking sectors only communicate via
gauge interactions which are common from the viewpoint of model building, it is argued
that the leading order contribution to the pseudo-goldstino mass arises at the three-loop level and
it could be described through the two-point correlation functions defined
in General Gauge Mediation~\cite{Meade:2008wd}.
Also a numerical value of the pseudo-goldstino mass was estimated from the evaluation of these functions
assuming the SUSY breaking scales $M_i$ in the two sectors are equal \cite{Argurio:2011hs}.
Note that the analysis in Ref.~\cite{Argurio:2011hs} is based on arguments and approximation,
while an independent cross-check or an explicit calculation of the three-loop contributions
with different messenger masses is still missing to date. Thus in this note we perform an
explicit three-loop calculation for the mass corrections to the pseudo-goldstino, which is
also applicable to the minimal gauge mediation.
Although the realistic SUSY breaking and gauge mediation models might not be as simple as the minimal
one mentioned above, the physics behind them for the mass corrections at the leading three-loop level
is almost the same. So our result could be easily generalized to other cases or used to make
model-dependent estimations.

This work is organized as follows.
In Section II we will make a brief review on the framework with pseudo-goldstinos and
mention some technical details used for our calculation. Section III contains the analytical
results and related discussions. Finally, we conclude in Section IV.

\section{Technical details}
\begin{figure}[t]
\begin{center}
\begin{tabular}{ccc}
\includegraphics[scale=0.3]{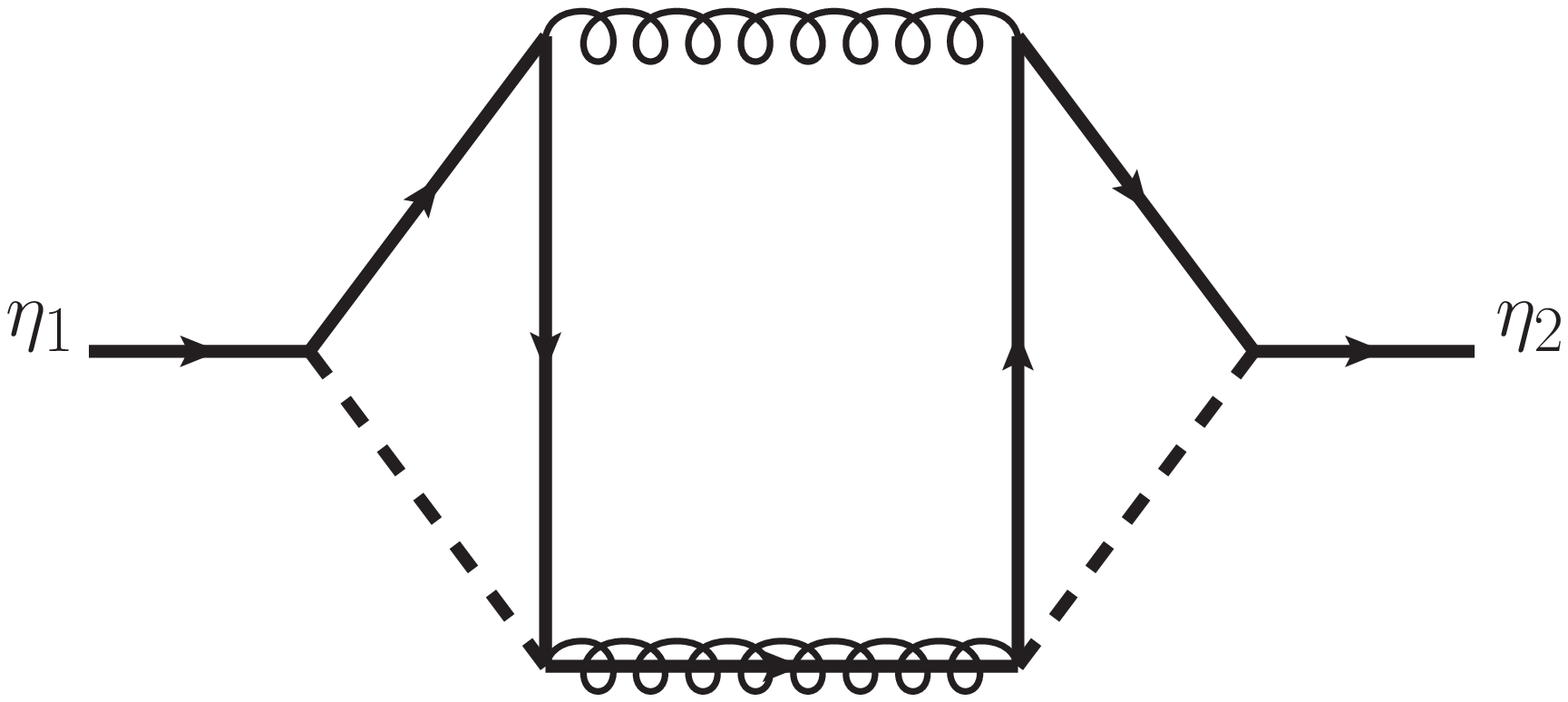}&
\hspace*{0mm}\includegraphics[scale=0.3]{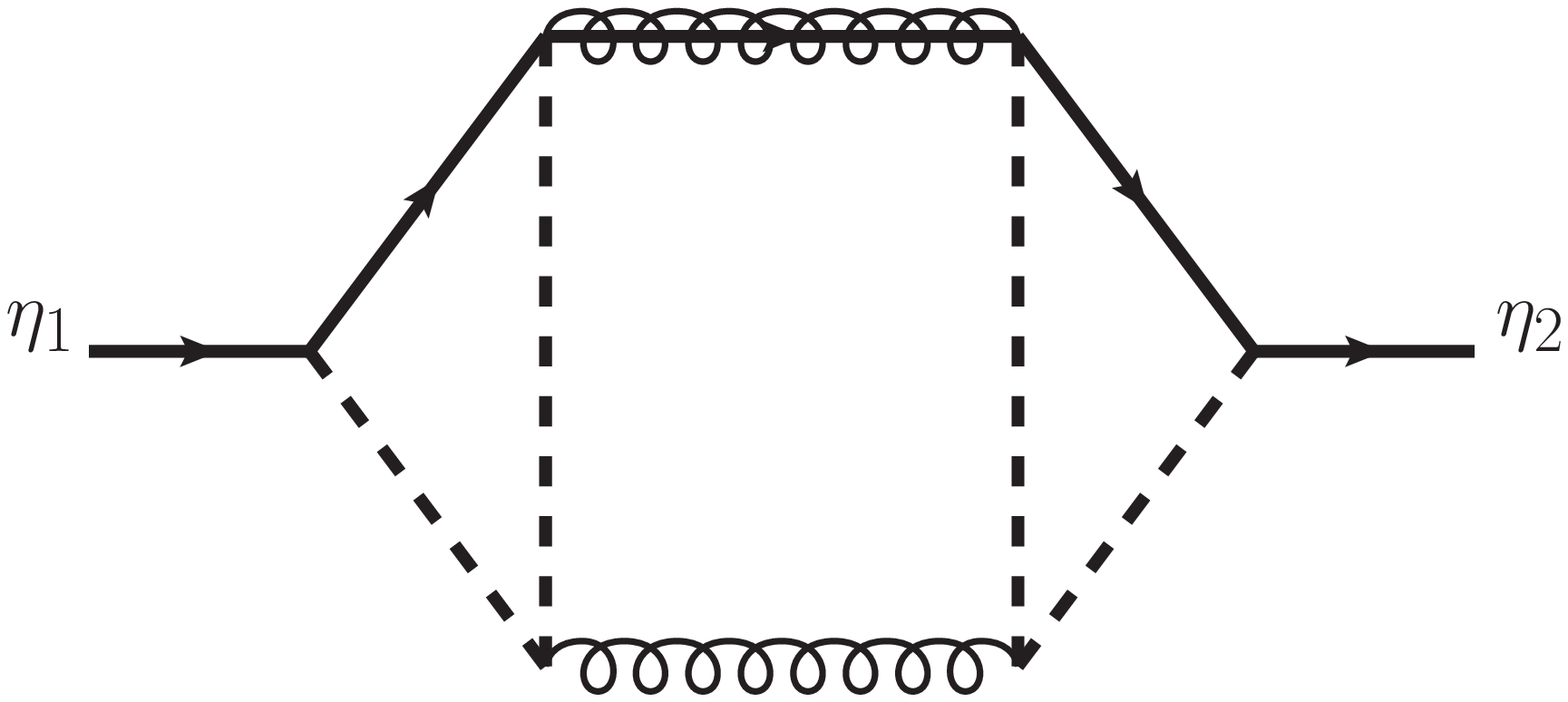}&
\hspace*{0mm}\includegraphics[scale=0.3]{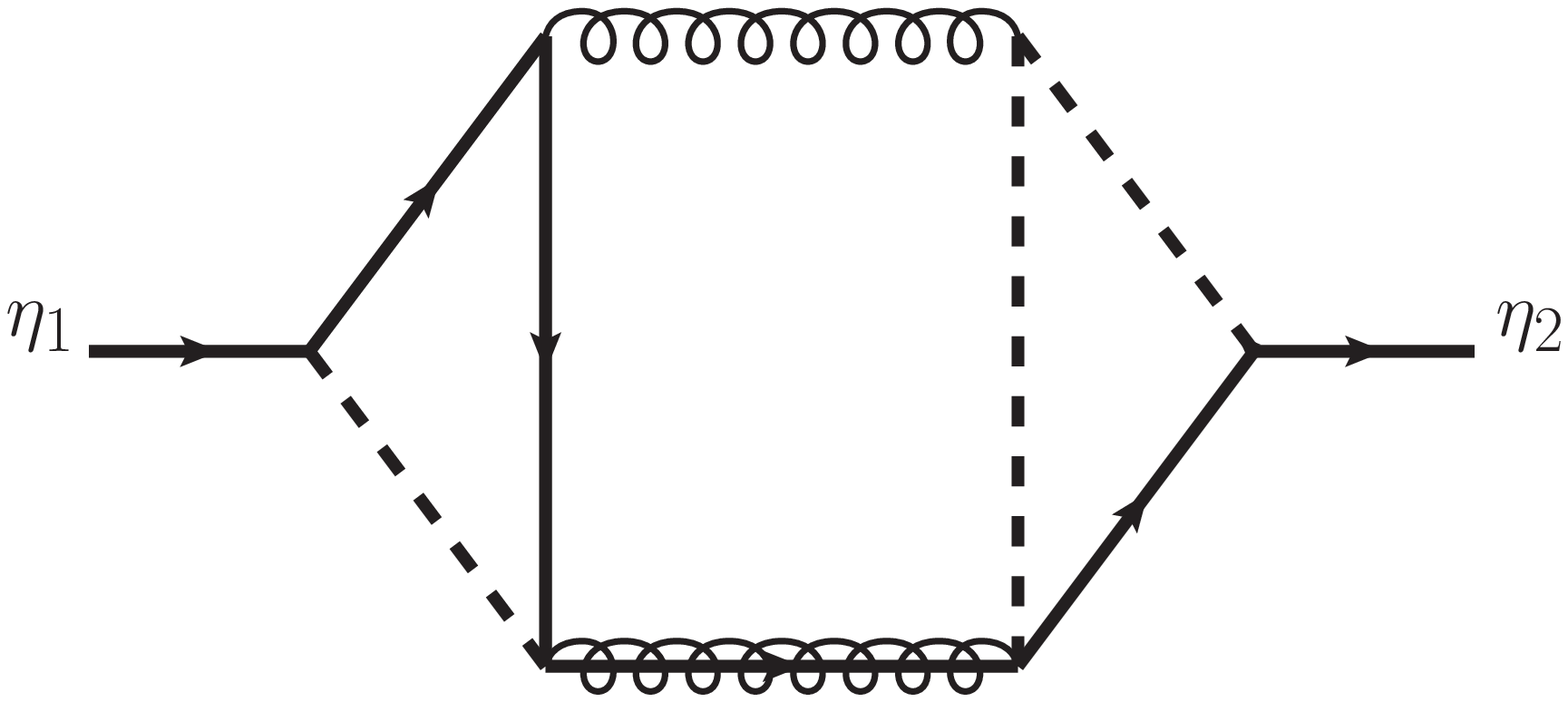}\\
(a)&\hspace*{0mm}(b)&\hspace*{0mm}(c)\\
\\
\includegraphics[scale=0.3]{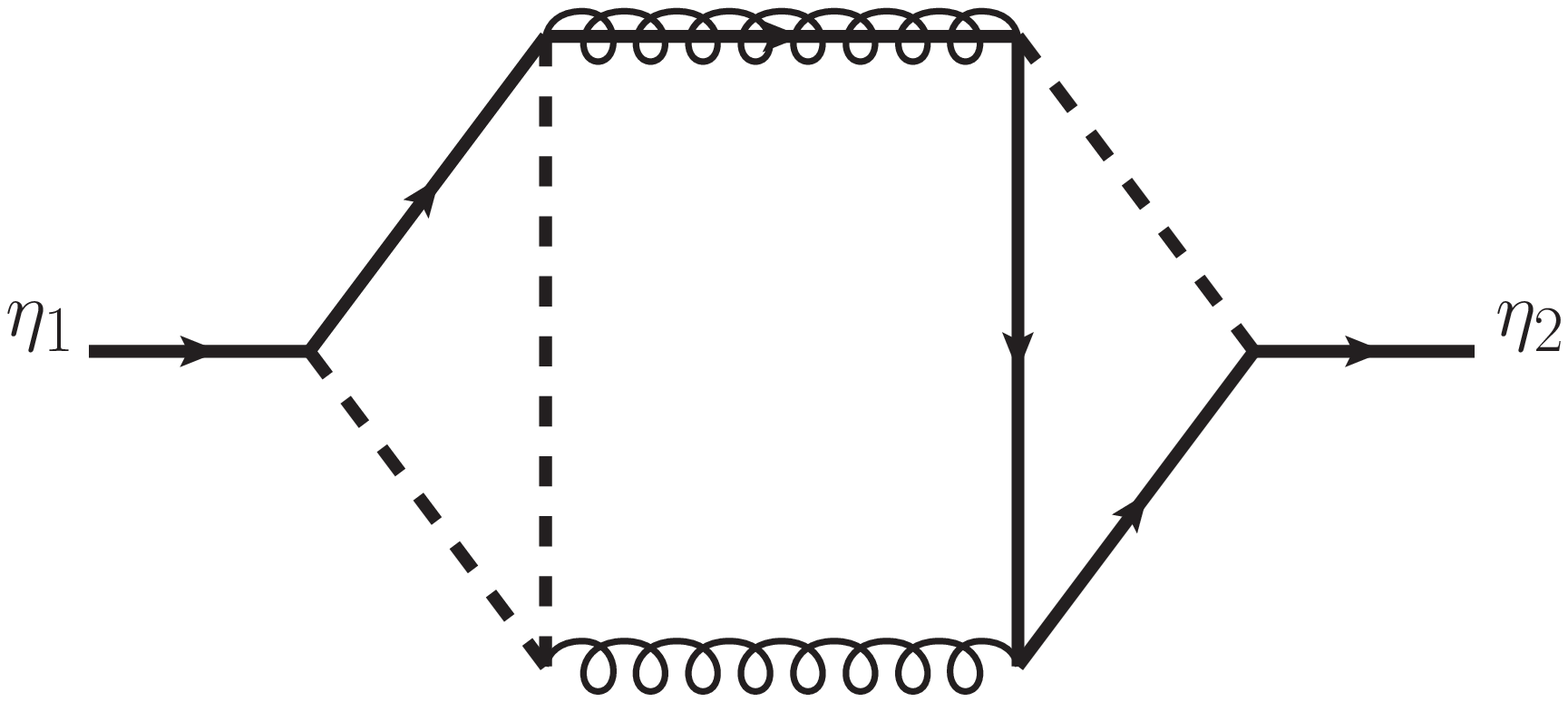}&
\hspace*{0mm}\includegraphics[scale=0.3]{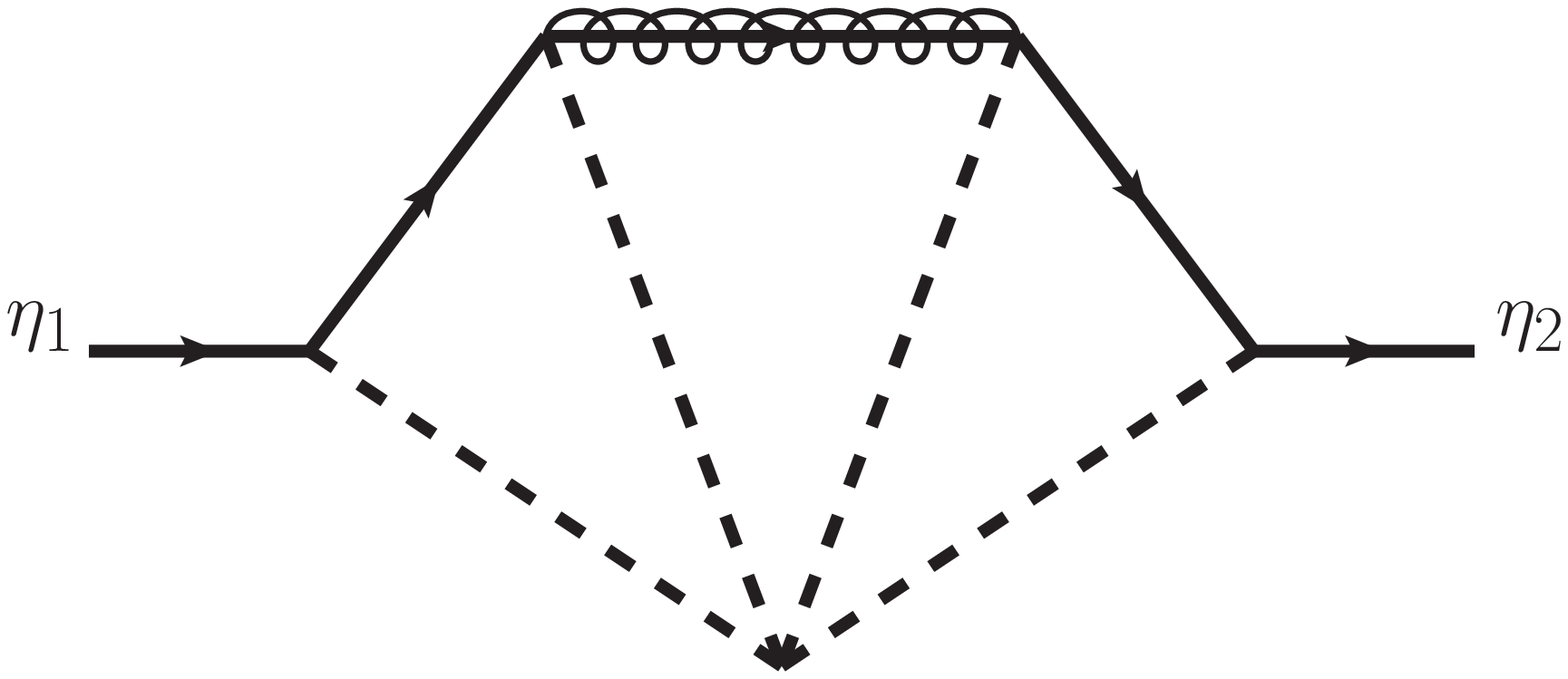}
\\
(d)&\hspace*{0mm}(e)\\
\end{tabular}
\end{center}
\caption{\label{fig::3loop}  Typical three-loop Feynman diagrams contributing to
the mass matrix element ${\cal M}_{12}$. The two sectors are connected via intermediate
gauge boson and its supersymmetric partner.}
\end{figure}

The superpotential with multi-sector SUSY breaking in the minimal gauge mediation is given by
\begin{align}
 W=X_i \Phi_i \bar\Phi_i + M_i \Phi_i \bar\Phi_i
 \label{eq::minimal}
\end{align}
with goldstino $\eta_i$ contained in the $\theta$ component of chiral spurion superfield $X_i$,
whose auxiliary component acquires F-term vacuum expectation value $F_i$.
To get more general results, here the messenger fields $\Phi_i$ and $\bar\Phi_i$
are assumed to fill the $N + {\bar N}$ representation of a general $SU(N)$ gauge group.
One could get similar corrections for $U(1)$ gauge transformations through replacing the overall $SU(N)$
color factor $C_N$ by $1$  after exchanging the gauge couplings
(the rules might not be so simple when going to higher loops).
The fermionic part of the messengers obtain Dirac masses $M_i$ which are not affected
by SUSY breaking. And the bosonic parts have the mass matrix
\begin{align}
	\begin{pmatrix}
	|M_i|^2  & \quad   F_i \\
	F_i^*    & \quad  |M_i|^2
	\end{pmatrix}
	,
\end{align}
whose two eigenvalues are $|M_i|^2 \pm F_i$. In order to avoid tachyonic scalar masses and gauge
symmetry breaking, usually it is assumed that $F_i \ll |M_i|^2$.

Next, let us turn to the goldstino part in Eq.~\ref{eq::minimal}.
In the two hidden sector scenario we define
$F=\sqrt{F_1^2+F_2^2}$ and $\tan\theta=F_2/F_1$, then the combination
$G= \eta_1 \cos \theta + \eta_2 \sin \theta$ is eaten by the super-Higgs mechanism,
while one pseudo-goldstino $G^\prime=-\eta_1\sin\theta+\eta_2\cos\theta$ is left.
Due to the fact that $G$ has to be massless before the gravitational effects are taken into account,
the radiatively generated mass matrix for goldstinos $\eta_{1,2}$ has to be of the following form
\begin{align}
	\begin{pmatrix}
 	-\frac{F_2}{F_1} {\cal M}_{12}  & \quad  {\cal M}_{12} \\
    {\cal M}_{12}         & \quad  -\frac{F_1}{F_2} {\cal M}_{12}
	\end{pmatrix}
	,
\end{align}
where ${\cal M}_{12}$ denotes the loop induced mass mixing between the two 
goldstinos $\eta_{1,2}$, i.e., $-\frac{1}{2}\eta_1{\cal M}_{12}\eta_2$.
Now it is easy to get the Majorana mass of the pseudo-goldstino
\begin{align}
m_{G^\prime} =\left(\frac{F_1}{F_2} + \frac{F_2}{F_1}\right) {\cal M}_{12}.
\label{eq::mass}
\end{align}

The typical non-vanishing Feynman diagrams contributing to ${\cal M}_{12}$ which start at three loops
are shown in Fig.~\ref{fig::3loop}, where the sectors are connected by supersymmetric gauge
interactions. Physically only the zero external momentum could generate the mass correction,
in other words we have to calculate the three-loop vacuum Feynman integrals.
On the technical side, the analytical formulas written in terms of polylogarithms for general vacuum
integrals at two-loop level have been known for some time.
But the general results at three-loop level are still not available \cite{Fleischer:1999mp, Chetyrkin:1999qi, Schroder:2005va, Grigo:2012ji,Bekavac:2009gz, Freitas:2016zmy, Martin:2016bgz, vanderBij:2000cg, Faisst:2003px, Kant:2010tf}. 
In the case of hierarchical momenta and masses, 
asymptotic expansions (see, e.g., Refs.~\cite{Smirnov:2002pj,Seidensticker:1999bb}) 
could be applied to evaluate Feynman integrals. 
As mentioned above, the scalar mass of the messengers will be split to
$|M_i|^2 \pm F_i$ after SUSY breaking. So there are four different mass scales in the first
four diagrams and six scales in the last one of Fig.~\ref{fig::3loop}.
Although these multi-scale integrals can not be evaluated directly,
we can perform Taylor expansion on the scalar propagators to reduce the number
of scales appearing in the integral
\begin{align}
 \frac{1}{\ell^2 -(M_i^2 \pm F_i)} = \frac{1}{\ell^2 -M_i^2}
 \left[ 1+ \frac{\pm F_i}{\ell^2 -M_i^2}  + ...\right].
\end{align}
This trick has been used by one of us in calculating the next-to-next-to-leading order hadronic
correction to the muon anomalous magnetic momentum \cite{Kurz:2014wya}.
We also checked that in this way the same expanded results for the one-loop gaugino
mass corrections would be reproduced as in Ref.~\cite{Martin:1996zb}.
Thus, at the cost of increasing the number of integrals, we are left with two-scale vacuum
integrals which could be solved analytically.

In our calculation the algebraic part is done with the help of {\tt FORM} \cite{Kuipers:2012rf}.
{\tt FIRE} \cite{Smirnov:2014hma} which implements the Laporta algorithm \cite{Laporta:2001dd}
is used to reduce all the scalar integrals into a basis of master integrals (MI). The analytical
results for these MIs which require the introduction of harmonic polylogarithms (HPLs) $H(a_1,...,a_k;x)$
are already known \cite{Grigo:2012ji}. For simplicity, we show the simplest HPLs
and more details on their properties could be found in Refs.~\cite{Remiddi:1999ew,Maitre:2005uu,Maitre:2007kp}:
\begin{eqnarray}
 H(1;x)&=& -\log(1-x), \nonumber\\
 H(0;x)&=&  \log(x),   \nonumber\\
 H(-1;x)&=&  \log(1+x).
 \label{eq::HPLs}
\end{eqnarray}

\section{Results and discussions}
From Eq.~\ref{eq::mass} we learn that the mass of pseudo-goldstino could be easily derived from the radiative corrections to the matrix element ${\cal M}_{12}$.
In the following we choose to present ${\cal M}_{12}$ by grouping terms according to the power
of two independent SUSY breaking scales $F_{1,2}$:
 \begin{align}
{\cal M}_{12}=\frac{1}{4} \frac{g^4}{(16\pi^2)^3} C_N \Biggl[  A_1 \frac{F_1}{M_1}
+A_2 \frac{F_2}{M_1}
  +B_1 \frac{F_1^3}{M_1^5}
  +B_2 \frac{F_1^2 F_2}{M_1^5}
  +B_3 \frac{F_1 F_2^2}{M_1^5}
  +B_4 \frac{F_2^3}{M_1^5}  \nonumber \\
+C_1 \frac{F_1^5 }{M_1^9}
+C_2 \frac{F_1^4 F_2}{M_1^9}
+C_3 \frac{F_1^3 F_2^2}{M_1^9}
+C_4 \frac{F_1^2 F_2^3}{M_1^9}
+C_5 \frac{F_1^1 F_2^4}{M_1^9}
+C_6 \frac{ F_2^5}{M_1^9} + ...   \Biggl],
\label{eq::expansion}
\end{align}
where $C_N=(N^2-1)/4$ is the color factor for $SU(N)$ gauge group and $g$ is
the corresponding gauge coupling. One new dimensionless variable $x=M_2/M_1$ is
introduced to eliminate the second messenger mass scale $M_2$,
so the coefficients $A_i$, $B_i$ and $C_i$ in the above equation
are just functions of $x$. Although ${\cal O}{(F_i^7)}$ terms with coefficients $D_i$
are too long to be shown, their contributions are considered in the following analysis.
The expressions of coefficients $A_i$, $B_i$ and $C_i$ are given by
   \begin{eqnarray}
   A_1&=&\dfrac{8}{(-1+x^2)}\Biggl[2xH(0,0;x)+(-1+x^2) \Big(H(-1,0,0;x)+H(1,0,0;x) \Big)\Biggl],
\\
\nonumber\\
   A_2&=& \dfrac{8}{x(-1+x^2)}\Biggl[ -2xH(0,0;x)+(-1+x^2)\Big(H(-1,0,0;x)+H(1,0,0;x) \Big) \Biggl],
   \end{eqnarray}
\begin{eqnarray}
B_1&=&\dfrac{1}{48x^5 (-1 + x^2)^3}
\Big[ -6 x^4 + 22 x^6 - 26 x^8 + 10 x^{10} + 6 x^4H(0;x) 
   \nonumber \\  
        & &\hspace*{4mm}
  -2 x^6H(0;x) + 2 x^8H(0;x) - 6 x^{10}H(0;x) - 
  6 x^4 H(0,0;x) 
     \nonumber \\  
        & &\hspace*{4mm}
  + 22 x^6 H(0,0;x) + 
  22 x^8 H(0,0;x) - 6 x^{10} H(0,0;x) + 
  3 x^3H(-1,0,0;x) 
     \nonumber \\  
        & &\hspace*{4mm}
  - 12 x^5H(-1,0,0;x) + 
  18 x^7H(-1,0,0;x) - 12 x^9H(-1,0,0;x)
  \nonumber \\  
        & &\hspace*{4mm}
   + 
  3 x^{11}H(-1,0,0;x) + 3 x^3 H(1,0,0;x)
   - 
  12 x^5 H(1,0,0;x)
  \nonumber \\  
        & &\hspace*{4mm}
   + 18 x^7 H(1,0,0;x) - 
  12 x^9 H(1,0,0;x) + 3 x^{11} H(1,0,0;x) \Big],
\end{eqnarray}    
      \begin{eqnarray}
      B_2&=&\dfrac{1}{48 x^5 (-1 + x^2)^3}
      \Big[-12 x^3 + 28  x^5 - 20  x^7 + 4 x^9 + 12  x^3 H(0;x)
      \nonumber \\
      & &\hspace*{4mm}
      - 116  x^5 H(0;x) + 164  x^7 H(0;x) - 60  x^9 H(0;x) -  12 x^3H(0,0; x)
      \nonumber \\
      & &\hspace*{4mm}
      -164 x^5 H(0,0; x) + 172  x^7H(0,0; x) - 60 x^9H(0,0; x) + 6 x^2 H(-1,0,0; x)
      \nonumber \\
      & &\hspace*{4mm}
       - 48  x^4 H(-1,0,0; x) + 108 x^6 H(-1,0,0; x) - 96  x^8 H(-1,0,0; x)
       \nonumber \\
       & &\hspace*{4mm}
       + 30  x^{10} H(-1,0,0; x) + 6  x^2 H(1,0,0; x) - 48 x^4 H(1,0,0; x)
       \nonumber \\
       & &\hspace*{4mm}
       + 108 x^6 H(1,0,0; x)-96 x^8 H(1,0,0; x) + 30 x^{10} H(1,0,0; x) \Big],
      \end{eqnarray}
%
        \begin{eqnarray}
        B_3&=&\dfrac{1}{48 x^5 (-1 + x^2)^3}\Big[-4  x^2 + 20 x^4 - 28 x^6 + 12 x^8
         - 60 x^2 H(0;x) + 164 x^4 H(0;x)
         \nonumber \\
         & &\hspace*{4mm}
        - 116  x^6 H(0;x) + 12 x^8 H(0;x) + 60 x^2H(0,0; x) - 172 x^4H(0,0; x)
         \nonumber \\
         & &\hspace*{4mm}
        + 164 x^6H(0,0; x) + 12 x^8H(0,0; x) - 30 x H(-1,0,0; x) + 96  x^3 H(-1,0,0; x)
         \nonumber \\
         & &\hspace*{4mm}
         - 108 x^5 H(-1,0,0; x) + 48 x^7 H(-1,0,0; x)
         - 6 x^9 H(-1,0,0; x)
         \nonumber \\
         & &\hspace*{4mm}
         - 30 x H(1,0,0; x) + 96 x^3 H(1,0,0; x)
         - 108 x^5 H(1,0,0; x)
          \nonumber \\
          & &\hspace*{4mm}
         + 48 x^7 H(1,0,0; x) - 6 x^9 H(1,0,0; x) \Big],
        \end{eqnarray}

\begin{eqnarray}
B_4&=&\dfrac{1}{48x^5 (-1 + x^2)^3}
\Big[ -10 x + 26 x^3 - 22 x^5 + 6 x^7 - 6 x H(0;x) + 
  2 x^3 H(0;x)
  \nonumber \\  
        & &\hspace*{4mm}
   - 2 x^5 H(0;x) + 6 x^7 H(0;x) + 
  6 xH(0,0;x) - 22 x^3H(0,0;x) - 
  22 x^5H(0,0;x)
  \nonumber \\  
        & &\hspace*{4mm}
   + 6 x^7H(0,0;x) - 
  3H(-1,0,0;x) + 12 x^2H(-1,0,0;x) - 
  18 x^4H(-1,0,0;x)
  \nonumber \\  
        & &\hspace*{4mm}
   + 12 x^6H(-1,0,0;x) - 
  3 x^8H(-1,0,0;x) - 3 H(1,0,0;x) + 
  12 x^2 H(1,0,0;x)
  \nonumber \\  
        & &\hspace*{4mm}
   - 18 x^4 H(1,0,0;x) + 
  12 x^6 H(1,0,0;x) - 3 x^8 H(1,0,0;x) \Big],
\end{eqnarray}

\begin{eqnarray}
C_1&=& \dfrac{1}{92160x^9 (-1 + x^2)^5}
\Big[ 2430 x^6 - 16560 x^8 + 42062 x^{10} - 51536 x^{12} + 32154 x^{14} 
\nonumber \\  
        & &\hspace*{4mm}
- 9728 x^{16} + 1178 x^{18} - 2430 x^6 H(0;x) + 
  16380 x^8 H(0;x) - 66602 x^{10} H(0;x) 
  \nonumber \\  
        & &\hspace*{4mm}
  + 
  133224 x^{12} H(0;x) - 127026 x^{14} H(0;x) + 
  56540 x^{16} H(0;x) - 10086 x^{18} H(0;x) 
  \nonumber \\  
        & &\hspace*{4mm}
  +2430 x^6H(0,0;x) - 15840 x^8H(0,0;x) + 
  27246 x^{10}H(0,0;x) - 59280 x^{12}H(0,0;x) 
  \nonumber \\  
        & &\hspace*{4mm}
  +100218 x^{14}H(0,0;x) - 63120 x^{16}H(0,0;x) + 
  14490 x^{18}H(0,0;x) 
  \nonumber \\  
        & &\hspace*{4mm}
  - 1215 x^5H(-1,0,0;x) + 
  8325 x^7H(-1,0,0;x) - 16155 x^9H(-1,0,0;x)
  \nonumber \\  
        & &\hspace*{4mm}
  -1575 x^{11}H(-1,0,0;x) + 43875 x^{13}H(-1,0,0;x) - 
  59985 x^{15}H(-1,0,0;x) 
  \nonumber \\  
        & &\hspace*{4mm}
  + 33975 x^{17}H(-1,0,0;x) - 
  7245 x^{19}H(-1,0,0;x) - 1215 x^5 H(1,0,0;x) 
  \nonumber \\  
        & &\hspace*{4mm}
  +8325 x^7 H(1,0,0;x) - 16155 x^9 H(1,0,0;x) - 
  1575 x^{11} H(1,0,0;x)
  \nonumber \\  
        & &\hspace*{4mm}
   + 43875 x^{13} H(1,0,0;x) - 
  59985 x^{15} H(1,0,0;x) + 33975 x^{17} H(1,0,0;x)
  \nonumber \\  
        & &\hspace*{4mm}
   - 
  7245 x^{19} H(1,0,0;x) \Big],
\end{eqnarray}

\begin{eqnarray}
C_2&=& \dfrac{1}{92160x^9 (-1 + x^2)^5}
\Big[ -2430 x^5 + 5760 x^7 - 9342 x^9 + 18288 x^{11} - 22266 x^{13} 
\nonumber \\  
        & &\hspace*{4mm} 
 +12816 x^{15} - 2826 x^{17} + 2430 x^5 H(0;x) - 
  5580 x^7 H(0;x) - 93318 x^9 H(0;x) 
  \nonumber \\  
        & &\hspace*{4mm}
  +293112 x^{11} H(0;x) - 345006 x^{13} H(0;x) + 
  188052 x^{15} H(0;x) - 39690 x^{17} H(0;x) 
  \nonumber \\  
        & &\hspace*{4mm}
  -2430 x^5H(0,0;x) + 5040 x^7H(0,0;x) - 
  148254 x^9H(0,0;x) + 341040 x^{11}H(0,0;x) 
  \nonumber \\  
        & &\hspace*{4mm}
  -365658 x^{13}H(0,0;x) + 191520 x^{15}H(0,0;x) - 
  39690 x^{17}H(0,0;x)  
  \nonumber \\  
        & &\hspace*{4mm}
  +1215 x^4H(-1,0,0;x)-2925 x^6H(-1,0,0;x) - 23445 x^8H(-1,0,0;x) 
  \nonumber \\  
        & &\hspace*{4mm}
  +118575 x^{10}H(-1,0,0;x) - 223875 x^{12}H(-1,0,0;x) + 
  212985 x^{14}H(-1,0,0;x)   
  \nonumber \\  
        & &\hspace*{4mm}
 - 102375 x^{16}H(-1,0,0;x) + 19845 x^{18}H(-1,0,0;x)
  + 1215 x^4 H(1,0,0;x)  
   \nonumber \\  
        & &\hspace*{4mm}
  -2925 x^6 H(1,0,0;x) - 23445 x^8 H(1,0,0;x)+ 
  118575 x^{10} H(1,0,0;x) 
   \nonumber \\  
        & &\hspace*{4mm}
  -223875 x^{12} H(1,0,0;x) + 
  212985 x^{14} H(1,0,0;x) - 102375 x^{16} H(1,0,0;x) 
   \nonumber \\  
        & &\hspace*{4mm}
 + 19845 x^{18} H(1,0,0;x) \Big],
\end{eqnarray}

\begin{eqnarray}
C_3&=& \dfrac{1}{92160x^9 (-1 + x^2)^5}
\Big[ -3150 x^4 + 13120 x^6 - 18190 x^8 + 9840 x^{10} - 490 x^{12} 
 \nonumber \\  
        & &\hspace*{4mm}
  -2480 x^{14} + 1350 x^{16} + 3150 x^4 H(0;x) - 
  16300 x^6 H(0;x) + 31530 x^8 H(0;x) 
   \nonumber \\  
        & &\hspace*{4mm}
  -16200 x^{10} H(0;x) + 3970 x^{12} H(0;x) - 
  7500 x^{14} H(0;x) + 1350 x^{16} H(0;x) 
   \nonumber \\  
        & &\hspace*{4mm}
  -3150 x^4H(0,0;x) + 15600 x^6H(0,0;x) - 
  32430 x^8H(0,0;x) + 37680 x^{10}H(0,0;x) 
   \nonumber \\  
        & &\hspace*{4mm}
 +18870 x^{12}H(0,0;x) - 7200 x^{14}H(0,0;x) + 
  1350 x^{16}H(0,0;x) 
   \nonumber \\  
        & &\hspace*{4mm}
   + 1575 x^3H(-1,0,0;x)- 
  8325 x^5H(-1,0,0;x) + 18675 x^7H(-1,0,0;x)
   \nonumber \\  
        & &\hspace*{4mm}
   - 
  23625 x^9H(-1,0,0;x) +19125 x^{11}H(-1,0,0;x) - 
  10575 x^{13}H(-1,0,0;x)  
   \nonumber \\  
        & &\hspace*{4mm}
  + 3825 x^{15}H(-1,0,0;x)-675 x^{17}H(-1,0,0;x) + 1575 x^3 H(1,0,0;x)
   \nonumber \\  
        & &\hspace*{4mm}
    - 
  8325 x^5 H(1,0,0;x) + 18675 x^7 H(1,0,0;x)- 
  23625 x^9 H(1,0,0;x) 
   \nonumber \\  
        & &\hspace*{4mm}
  + 19125 x^{11} H(1,0,0;x) - 
  10575 x^{13} H(1,0,0;x) +3825 x^{15} H(1,0,0;x)
     \nonumber \\  
        & &\hspace*{4mm}
   - 
  675 x^{17} H(1,0,0;x) \Big], 
\end{eqnarray}

\begin{eqnarray}
C_4&=& \dfrac{1}{92160x^9 (-1 + x^2)^5}
\Big[ -1350 x^3 + 2480 x^5 + 490 x^7 - 9840 x^9 + 18190 x^{11} 
 \nonumber \\  
        & &\hspace*{4mm}
  -13120 x^{13} + 3150 x^{15} + 1350 x^3 H(0;x) - 
  7500 x^5 H(0;x) + 3970 x^7 H(0;x)
 \nonumber \\  
        & &\hspace*{4mm}
  -16200 x^9 H(0;x) + 31530 x^{11} H(0;x) - 
  16300 x^{13} H(0;x) + 3150 x^{15} H(0;x) 
   \nonumber \\  
        & &\hspace*{4mm}
 -1350 x^3H(0,0;x) + 7200 x^5H(0,0;x) - 
  18870 x^7H(0,0;x) - 37680 x^9H(0,0;x) 
   \nonumber \\  
        & &\hspace*{4mm}
  +32430 x^{11}H(0,0;x) - 15600 x^{13}H(0,0;x) + 
  3150 x^{15}H(0,0;x) + 675 x^2H(-1,0,0;x) 
   \nonumber \\  
        & &\hspace*{4mm}
 -3825 x^4H(-1,0,0;x) + 10575 x^6H(-1,0,0;x) - 
  19125 x^8H(-1,0,0;x) 
   \nonumber \\  
        & &\hspace*{4mm}
  +23625 x^{10}H(-1,0,0;x) - 
  18675 x^{12}H(-1,0,0;x) + 8325 x^{14}H(-1,0,0;x)
   \nonumber \\  
        & &\hspace*{4mm}
   -
  1575 x^{16}H(-1,0,0;x) + 675 x^2 H(1,0,0;x) - 
  3825 x^4 H(1,0,0;x) 
   \nonumber \\  
        & &\hspace*{4mm} 
  + 10575 x^6 H(1,0,0;x) -19125 x^8 H(1,0,0;x) + 23625 x^{10} H(1,0,0;x)
   \nonumber \\  
        & &\hspace*{4mm}
    - 18675 x^{12} H(1,0,0;x) + 8325 x^{14}H(1,0,0;x) - 
  1575 x^{16} H(1,0,0;x) \Big],
\end{eqnarray}

\begin{eqnarray}
C_5&=& -\dfrac{1}{92160x^9 (-1 + x^2)^5}
\Big[-2826  x^2 + 12816  x^4 - 22266  x^6 + 18288  x^8 - 9342  x^{10}
            \nonumber \\  
            & &\hspace*{4mm}
            + 5760  x^{12} - 2430  x^{14} + 
           39690  x^2 H(0;x) - 188052  x^4 H(0;x) + 
           345006 x^6 H(0;x) 
            \nonumber \\  
            & &\hspace*{4mm}
            - 293112  x^8 H(0;x) + 
           93318  x^{10} H(0;x) + 5580  x^{12} H(0;x) - 
           2430  x^{14} H(0;x)
            \nonumber \\  
            & &\hspace*{4mm}
            - 39690  x^2H(0,0; x) + 
           191520  x^4H(0,0; x) - 365658  x^6H(0,0; x)  
            \nonumber \\  
            & &\hspace*{4mm}
           + 341040 x^8H(0,0; x) - 148254  x^{10}H(0,0; x) + 5040  x^{12}H(0,0; x)  
            \nonumber \\  
            & &\hspace*{4mm}
            - 2430  x^{14}H(0,0; x)+ 19845 x H(-1,0,0; x) - 
           102375  x^3 H(-1,0,0; x) 
            \nonumber \\  
            & &\hspace*{4mm}
            + 
           212985  x^5 H(-1,0,0; x)- 223875  x^7 H(-1,0,0; x) + 
           118575  x^9 H(-1,0,0; x)
            \nonumber \\  
            & &\hspace*{4mm}
             - 
           23445  x^{11} H(-1,0,0; x)- 2925  x^{13} H(-1,0,0; x) + 
           1215  x^{15} H(-1,0,0; x) 
            \nonumber \\  
            & &\hspace*{4mm}
           + 
           19845  x H(1,0,0; x)  - 102375  x^3 H(1,0,0; x) +212985  x^5 H(1,0,0; x)
            \nonumber \\  
            & &\hspace*{4mm}
            - 
           223875  x^7 H(1,0,0; x) + 118575  x^9 H(1,0,0; x) - 
           23445  x^{11} H(1,0,0; x) 
           \nonumber \\  
            & &\hspace*{4mm}
           - 
           2925  x^{13} H(1,0,0; x) + 1215  x^{15} H(1,0,0; x) \Big], 
\end{eqnarray}

\begin{eqnarray}
C_6&=& \dfrac{1}{92160x^9 (-1 + x^2)^5}
\Big[ -1178 x + 9728 x^3 - 32154 x^5 + 51536 x^7 - 42062 x^9 
 \nonumber \\  
        & &\hspace*{4mm}
  +16560 x^{11} - 2430 x^{13} - 10086 x H(0;x) + 
  56540 x^3 H(0;x) - 127026 x^5 H(0;x)  
   \nonumber \\  
        & &\hspace*{4mm}
  +133224 x^7 H(0;x) - 66602 x^9 H(0;x) + 
  16380 x^{11} H(0;x) - 2430 x^{13} H(0;x) 
   \nonumber \\  
        & &\hspace*{4mm}
  -14490 xH(0,0;x) + 63120 x^3H(0,0;x) - 
  100218 x^5H(0,0;x) + 59280 x^7H(0,0;x)  
   \nonumber \\  
        & &\hspace*{4mm}
  -27246 x^9H(0,0;x) + 15840 x^{11}H(0,0;x) - 
  2430 x^{13}H(0,0;x) + 7245H(-1,0,0;x) 
   \nonumber \\  
        & &\hspace*{4mm}
  -33975x^2H(-1,0,0;x) + 59985 x^4H(-1,0,0;x) - 
  43875 x^6H(-1,0,0;x)  
   \nonumber \\  
        & &\hspace*{4mm}
  + 1575 x^8H(-1,0,0;x) +16155 x^{10}H(-1,0,0;x) - 8325 x^{12}H(-1,0,0;x) 
   \nonumber \\  
        & &\hspace*{4mm}
 + 
  1215 x^{14}H(-1,0,0;x) + 7245 H(1,0,0;x)  -33975 x^2 H(1,0,0;x) 
   \nonumber \\  
        & &\hspace*{4mm}
 + 59985 x^4 H(1,0,0;x) - 
  43875 x^6 H(1,0,0;x) + 1575 x^8 H(1,0,0;x) 
   \nonumber \\  
        & &\hspace*{4mm}  
   +16155 x^{10} H(1,0,0;x) - 8325 x^{12} H(1,0,0;x) + 
  1215 x^{14} H(1,0,0;x) \Big]. 
\end{eqnarray}

From above equations, it is easy to see that there are no contributions with even powers of ${F_i}$.
The reason is rather simple: perturbative expansion of ${F_i}$ in the Feynman diagrams
plays the role of converting the scalar components of the superfields from one to another,
and even times of these transformations lead to SUSY-preserving interactions which should
vanish due to the symmetry constraint. Same behaviors have also been found for the one-loop
gaugino mass corrections in the model of minimal gauge mediation \cite{Martin:1996zb},
where even powers of $F_i$ also do not contribute.

\begin{table}[t]
\begin{center}
    \begin{tabular}{|c|c|c|c|c|}
     \hline
     & $i=1$ & $i=2$ & $i=3$  & $i=4$  \\
      \hline
      \hline
 ~$A_i$~ &$14\zeta_3$ & & & \\ \hline
 ~$B_i$~ &$\frac{1}{6} $
         & $-\frac{1}{12} + \frac{7}{8}\zeta_3$ & & \\ \hline
 ~$C_i$~ & ~$-\frac{61}{4608} - \frac{161}{1024}\zeta_3$~
         & ~$-\frac{77}{1536} + \frac{301}{2024}\zeta_3$~
         & ~$\frac{265}{4608} - \frac{35}{1024}\zeta_3$~  & \\ \hline
 ~$D_i$~ & ~$-\frac{21727}{1105920}-\frac{2149}{16384}\zeta_3~$
         & ~$-\frac{1411}{46080}+\frac{301}{2048}\zeta_3$~
         & ~$\frac{1}{60}$~
         & ~$\frac{1663}{61440}-\frac{147}{8192}\zeta_3$~\\
       \hline
    \end{tabular}
\end{center}
    \caption{\label{tab::1}
Values of $A_i$, $B_i$, $C_i$ and $D_i$ defined in Eq.~(\ref{eq::expansion})
when the two messenger masses are equal. Symmetric parts are not displayed.
$\zeta_n$ denotes the Riemann's zeta function.
}
\end{table}

Next, another important check besides UV-finiteness\footnote{Here no renormalization is needed since the first non-vanishing correction starts at three loops. 
All the calculations are performed in four dimensions before obtaining the finite intermediate results which 
are expressed in terms of scalar integrals. So, there should be no violations of SUSY relations 
as discussed, e.g., in Ref.~\cite{Stockinger:2018oxe}}
on the correctness of our result will be explained. When $x$ is fixed to be 1,
the result for the related functions expressed in terms of complicated
HPLs could be greatly simplified. In this case, one will get the expressions
listed in Table~\ref{tab::1}. It means that when the two messenger masses are the same,
i.e., we can set $M_1=M_2$ at the beginning of the calculation, then we do the reduction
for single-scale integrals and obtain single-scale MIs. 
The second approach has been done independently and thus provides a good cross-check.

\begin{figure}[t]
\begin{center}
\begin{tabular}{cc}
\includegraphics[scale=0.8]{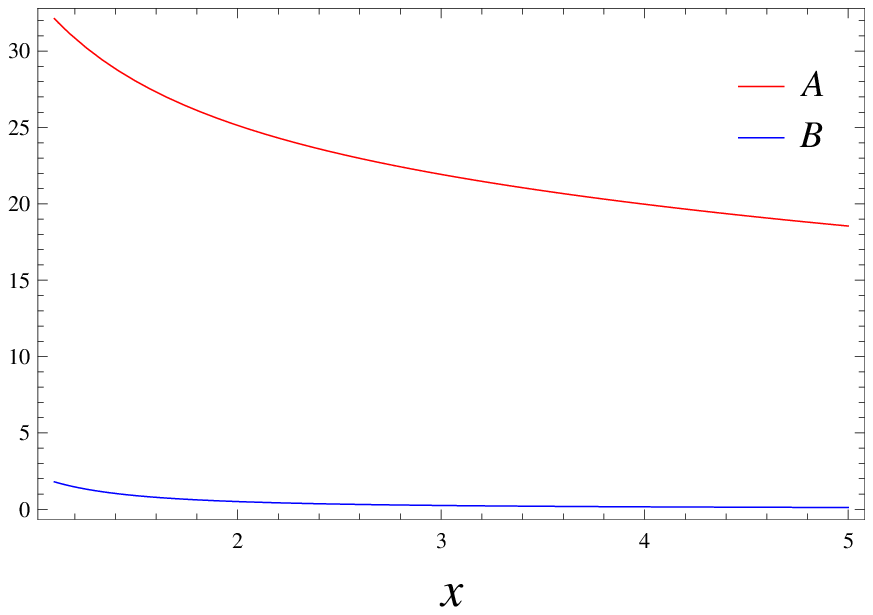}&
\includegraphics[scale=0.8]{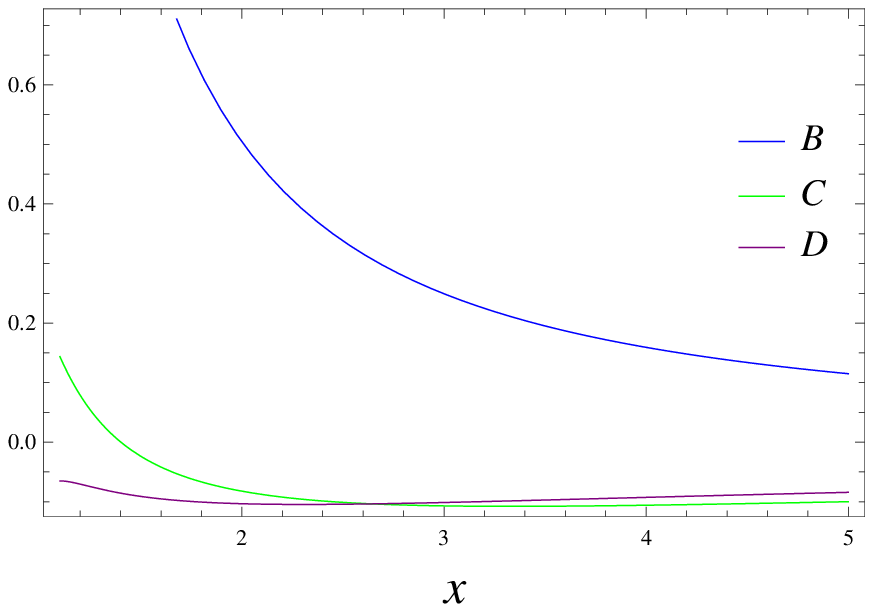}
\\
(a)&(b)
\end{tabular}
\end{center}
\caption{\label{fig::plot} Plots of the functions $A$, $B$, $C$ and $D$.}
\end{figure}

With the assumption $F_A=F_B$, at the leading power of $F_i$ one could define a new
function $A=\sum\limits_{i} A_i$ to parameterize the radiative corrections.
From the arguments based on internal symmetry $M_1\leftrightarrow M_2$,
we know that $A(x)$ should equal to $ A(1/x)/x$ which has been verified.
Similar self-consistent checks have also been performed for new functions $B$, $C$ and $D$.
In order to get an estimation of the numerical effects of the higher order terms expanded in $F_i$,
two plots are shown in Fig.~\ref{fig::plot}. Obviously, the functions $B$, $C$, $D$ are much smaller
than $A$ for moderate values of $x$. The conclusion does not change for large values of $x$, e.g.,
we get $A\simeq 4.30$, $B\simeq0.003$, $C\simeq-0.01$, $D\simeq-0.01$ when $x=100$.
So, the higher order terms could be safely neglected. From Table \ref{tab::2}
and Fig.~\ref{fig::plot}(a) one could see that the value of $A$ decreases with the increase of $x$.
It means we can get a much lighter pseudo-goldstino when the messenger scales are different.

Note that comparing to the usual low energy soft SUSY breaking parameter $m_{soft}$, the pseudo-goldstino
mass is suppressed by a factor $g^2/(16\pi^2)^2$ when both the SUSY breaking and messenger scales are comparable.
Considering the summation effects over different gauge sectors, the authors in Ref.~\cite{Argurio:2011hs}
concluded that the mass of pseudo-goldstino is around the GeV scale in this case. Thus, it is rather simple
to see from Table ~\ref{tab::2} that the pseudo-goldstino mass can be further suppressed by the ratio 
of messenger masses, e.g., as low as ${{\cal O} (0.1)}$ GeV when $x=M_2/M_1=100$.

\begin{table}[t]
\begin{center}
    \begin{tabular}{|c|c|c|c|c|c|}
     \hline
     & ~$x=1$~ & ~$x=10$~ & ~$x=10^2$~  & ~$x=10^3$~ & ~$x=10^4$~ \\
      \hline
      \hline
 ~$A$~ &$33.66$ & $14.36$ & $4.30$
       &$0.89$  & $0.15$\\ \hline
    \end{tabular}
\end{center}
    \caption{\label{tab::2}
Approximated values of $A$ for different ratios of messenger scales.}
\end{table}

Before the end of this section, let us make a direct comparison with the result obtained
in \cite{Argurio:2011hs}. The authors considered the scenario
of equal messenger scales and then performed numerical evaluation. The approximated
value $4.21$ in Appendix B of Ref.~\cite{Argurio:2011hs} agrees well with our analytical
result at ${\cal O} (F_i)$ which corresponding to $\frac{7}{2}\zeta_3$, when the same
convention as theirs is used.

\section{Conclusion}

We calculated the three-loop contribution to the mass of pseudo-goldstino explicitly.
Since the exact evaluation of a basis of MIs for general vacuum integrals at the three-loop level
is still unavailable, the implicit conditions $F_i \ll M_i^2$ are used to reduce the number of scales we have to deal with.
After performing expansions in $F_i$, we provided analytical results which are in good agreement
with the numerical value in the literature. It has been proved that the mass correction is dominated
by the leading order contribution at ${\cal O}(F_i)$. We also found that the mass correction
could be greatly reduced if there is a hierarchy between different messenger scales.
Note that although the whole calculation is based on the two-sector SUSY-breaking scenario,
our results could also be generalized to multi-sector cases as long as these sectors
communicate with each other via gauge interactions.
So, our result presented in this note could be useful for phenomenological
analysis or model-buildings on pseudo-goldstino in the future.

\section*{Acknowledgments}
This work was supported in part by IHEP under Grant No. Y9515570U1,
by the National Natural Science Foundation of China (NNSFC)
under grant Nos. 11821505 and 12075300, by Peng-Huan-Wu Theoretical Physics Innovation
Center (12047503), by the CAS Center for Excellence in Particle Physics (CCEPP), by the
CAS Key Research Program of Frontier Sciences, and by a Key R\&D Program of Ministry
of Science and Technology of China under number 2017YFA0402204.


\begin{thebibliography}{99}

\bibitem{Argurio:2011hs}
R.~Argurio, Z.~Komargodski and A.~Mariotti,
Phys. Rev. Lett. \textbf{107}, 061601 (2011)
[arXiv:1102.2386 [hep-th]].


\bibitem{Meade:2008wd}
P.~Meade, N.~Seiberg and D.~Shih,
Prog. Theor. Phys. Suppl. \textbf{177} (2009), 143-158
[arXiv:0801.3278 [hep-ph]].

\bibitem{Cheung:2010mc}
C.~Cheung, Y.~Nomura and J.~Thaler,
JHEP \textbf{03} (2010), 073
[arXiv:1002.1967 [hep-ph]].

\bibitem{Cheung:2010qf}
C.~Cheung, J.~Mardon, Y.~Nomura and J.~Thaler,
JHEP \textbf{07} (2010), 035
[arXiv:1004.4637 [hep-ph]].

\bibitem{Craig:2010yf}
N.~Craig, J.~March-Russell and M.~McCullough,
JHEP \textbf{10} (2010), 095
[arXiv:1007.1239 [hep-ph]].

\bibitem{McCullough:2010wf}
M.~McCullough,
Phys. Rev. D \textbf{82} (2010), 115016
[arXiv:1010.3203 [hep-ph]].

\bibitem{Cheng:2010mw}
H.~C.~Cheng, W.~C.~Huang, I.~Low and A.~Menon,
JHEP \textbf{03} (2011), 019
[arXiv:1012.5300 [hep-ph]].

\bibitem{Izawa:2011hi}
K.~I.~Izawa, Y.~Nakai and T.~Shimomura,
JHEP \textbf{03} (2011), 007
[arXiv:1101.4633 [hep-ph]].

\bibitem{Thaler:2011me}
J.~Thaler and Z.~Thomas,
JHEP \textbf{07} (2011), 060
[arXiv:1103.1631 [hep-ph]].

\bibitem{Cheung:2011jq}
C.~Cheung, F.~D'Eramo and J.~Thaler,
JHEP \textbf{08} (2011), 115
[arXiv:1104.2600 [hep-ph]].

\bibitem{Dudas:2011kt}
E.~Dudas, G.~von Gersdorff, D.~M.~Ghilencea, S.~Lavignac and J.~Parmentier,
Nucl. Phys. B \textbf{855} (2012), 570-591
[arXiv:1106.5792 [hep-th]].

\bibitem{Argurio:2011gu}
R.~Argurio, K.~De Causmaecker, G.~Ferretti, A.~Mariotti, K.~Mawatari and Y.~Takaesu,
JHEP \textbf{06} (2012), 096
[arXiv:1112.5058 [hep-ph]].

\bibitem{Liu:2013sx}
T.~Liu, L.~Wang and J.~M.~Yang,
Phys. Lett. B \textbf{726} (2013), 228-233
[arXiv:1301.5479 [hep-ph]].

\bibitem{Ferretti:2013wya}
G.~Ferretti, A.~Mariotti, K.~Mawatari and C.~Petersson,
JHEP \textbf{04} (2014), 126
[arXiv:1312.1698 [hep-ph]].

\bibitem{Hikasa:2014yra}
K.~i.~Hikasa, T.~Liu, L.~Wang and J.~M.~Yang,
JHEP \textbf{07} (2014), 065
[arXiv:1403.5731 [hep-ph]].

\bibitem{Liu:2014lda}
T.~Liu, L.~Wang and J.~M.~Yang,
JHEP \textbf{02} (2015), 177
[arXiv:1411.6105 [hep-ph]].

\bibitem{Chen:2021omv}
J.~Chen, C.~Han, J.~M.~Yang and M.~Zhang,
[arXiv:2101.12131 [hep-ph]].

\bibitem{Cao:2020oxq}
J.~Cao, X.~Du, Z.~Li, F.~Wang and Y.~Zhang,
[arXiv:2007.09981 [hep-ph]].

\bibitem{Wang:2021bcx}
F.~Wang, L.~Wu, Y.~Xiao, J.~M.~Yang and Y.~Zhang,
[arXiv:2104.03262 [hep-ph]].

\bibitem{Fleischer:1999mp}
J.~Fleischer and M.~Y.~Kalmykov,
Phys. Lett. B \textbf{470} (1999), 168-176
[arXiv:hep-ph/9910223 [hep-ph]].

\bibitem{Chetyrkin:1999qi}
K.~G.~Chetyrkin and M.~Steinhauser,
Nucl. Phys. B \textbf{573} (2000), 617-651
[arXiv:hep-ph/9911434 [hep-ph]].

\bibitem{Schroder:2005va}
Y.~Schroder and A.~Vuorinen,
JHEP \textbf{06} (2005), 051
[arXiv:hep-ph/0503209 [hep-ph]].

\bibitem{Grigo:2012ji}
J.~Grigo, J.~Hoff, P.~Marquard and M.~Steinhauser,
Nucl. Phys. B \textbf{864} (2012), 580-596
[arXiv:1206.3418 [hep-ph]].

\bibitem{Bekavac:2009gz}
S.~Bekavac, A.~G.~Grozin, D.~Seidel and V.~A.~Smirnov,
Nucl. Phys. B \textbf{819} (2009), 183-200
[arXiv:0903.4760 [hep-ph]].

\bibitem{Freitas:2016zmy}
A.~Freitas,
JHEP \textbf{11} (2016), 145
[arXiv:1609.09159 [hep-ph]].

\bibitem{Martin:2016bgz}
S.~P.~Martin and D.~G.~Robertson,
Phys. Rev. D \textbf{95} (2017) no.1, 016008
[arXiv:1610.07720 [hep-ph]].

\bibitem{vanderBij:2000cg}
J.~J.~van der Bij, K.~G.~Chetyrkin, M.~Faisst, G.~Jikia and T.~Seidensticker,
Phys. Lett. B \textbf{498} (2001), 156-162
[arXiv:hep-ph/0011373 [hep-ph]].

\bibitem{Faisst:2003px}
M.~Faisst, J.~H.~Kuhn, T.~Seidensticker and O.~Veretin,
Nucl. Phys. B \textbf{665} (2003), 649-662
[arXiv:hep-ph/0302275 [hep-ph]].

\bibitem{Kant:2010tf}
P.~Kant, R.~V.~Harlander, L.~Mihaila and M.~Steinhauser,
JHEP \textbf{08} (2010), 104
[arXiv:1005.5709 [hep-ph]].

\bibitem{Smirnov:2002pj}
V.~A.~Smirnov,
Springer Tracts Mod. Phys. \textbf{177} (2002), 1-262

\bibitem{Seidensticker:1999bb}
T.~Seidensticker,
[arXiv:hep-ph/9905298 [hep-ph]].

\bibitem{Kurz:2014wya}
A.~Kurz, T.~Liu, P.~Marquard and M.~Steinhauser,
Phys. Lett. B \textbf{734} (2014), 144-147
[arXiv:1403.6400 [hep-ph]].

\bibitem{Martin:1996zb}
S.~P.~Martin,
Phys. Rev. D \textbf{55} (1997), 3177-3187
[arXiv:hep-ph/9608224 [hep-ph]].

\bibitem{Kuipers:2012rf}
J.~Kuipers, T.~Ueda, J.~A.~M.~Vermaseren and J.~Vollinga,
Comput. Phys. Commun. \textbf{184} (2013), 1453-1467
[arXiv:1203.6543 [cs.SC]].

\bibitem{Smirnov:2014hma}
A.~V.~Smirnov,
Comput. Phys. Commun. \textbf{189} (2015), 182-191
[arXiv:1408.2372 [hep-ph]].

\bibitem{Laporta:2001dd}
S.~Laporta,
Int. J. Mod. Phys. A \textbf{15} (2000), 5087-5159
[arXiv:hep-ph/0102033 [hep-ph]].


\bibitem{Remiddi:1999ew}
E.~Remiddi and J.~A.~M.~Vermaseren,
Int. J. Mod. Phys. A \textbf{15} (2000), 725-754
[arXiv:hep-ph/9905237 [hep-ph]].

\bibitem{Maitre:2005uu}
D.~Maitre,
Comput. Phys. Commun. \textbf{174} (2006), 222-240
[arXiv:hep-ph/0507152 [hep-ph]].

\bibitem{Maitre:2007kp}
D.~Maitre,
Comput. Phys. Commun. \textbf{183} (2012), 846
[arXiv:hep-ph/0703052 [hep-ph]].

\bibitem{Stockinger:2018oxe}
D.~St\"ockinger and J.~Unger,
Nucl. Phys. B \textbf{935} (2018), 1-16
[arXiv:1804.05619 [hep-ph]].

\end{thebibliography}
\end{document}